\documentclass[9pt,twocolumn,twoside]{arxiv-styles/opticajnl}
\journal{arxiv} 
\usepackage{helvet} 

\newcommand{\bv}[1]{\mathbf{#1}}
\setboolean{shortarticle}{true}

\title{Single-shot volumetric fluorescence imaging with neural fields}

\author[1,\dag,*]{Oumeng Zhang}
\author[1,\dag]{Haowen Zhou}
\author[2]{Brandon Y. Feng}
\author[3]{Elin M. Larsson}
\author[3]{Reinaldo E. Alcalde}
\author[4]{Siyuan Yin}
\author[1]{Catherine Deng}
\author[1,4]{Changhuei Yang}

\affil[1]{Department of Electrical Engineering, California Institute of Technology, Pasadena, California 91125, USA}
\affil[2]{Computer Science and Artificial Intelligence Laboratory, Massachusetts Institute of Technology, Cambridge, MA 02139, USA}
\affil[3]{Division of Biology and Biological Engineering, California Institute of Technology, Pasadena, California 91125, USA}
\affil[4]{Department of Medical Engineering, California Institute of Technology, Pasadena, California 91125, USA}
\affil[$\dag$]{These authors contributed equally to this work.}
\affil[*]{ozhang@caltech.edu}

\begin{abstract}
Single-shot volumetric fluorescence (SVF) imaging offers a significant advantage over traditional imaging methods that require scanning across multiple axial planes as it can capture biological processes with high temporal resolution. 
The key challenges in SVF imaging include requiring sparsity constraints, eliminating depth ambiguity in the reconstruction, and maintaining high resolution across a large field of view.
In this paper, we introduce the QuadraPol point spread function (PSF) combined with neural fields, a novel approach for SVF imaging. 
This method utilizes a custom polarizer at the back focal plane and a polarization camera to detect fluorescence, effectively encoding the 3D scene within a compact PSF without depth ambiguity. 
Additionally, we propose a reconstruction algorithm based on the neural fields technique that provides improved reconstruction quality compared to classical deconvolution methods.
QuadraPol PSF, combined with neural fields, significantly reduces the acquisition time of a conventional fluorescence microscope by approximately 20 times and captures a 100~mm$^3$ cubic volume in one shot. 
We validate the effectiveness of both our hardware and algorithm through all-in-focus imaging of bacterial colonies on sand surfaces and visualization of plant root morphology. 
Our approach offers a powerful tool for advancing biological research and ecological studies.
\end{abstract}

\setboolean{displaycopyright}{false} 

\begin{document}

\maketitle

\section{Introduction}

Fluorescence imaging is an indispensable tool in biological research as it enables real-time observation of live organisms due to its high sensitivity, biological specificity, and non-invasive nature. 
However, conventional 2D fluorescence microscopy cannot capture the complete 3D structure of biological samples. 
To address this, techniques such as confocal and light sheet fluorescence microscopy have been developed and are widely used.
Despite their advantages, these methods require scanning multiple axial planes, significantly increasing the acquisition time and limiting the spatial-temporal throughput.
The extensive scanning time is impractical for numerous applications where imaging large fields of view (FOVs) is required, such as when performing lab studies of the complex biological processes associated with the rhizosphere \cite{prashar_rhizosphere_2014} - the media-root interfaces of plants.

Single-shot volumetric fluorescence (SVF) imaging techniques have been developed to address the challenges in scanning-based 3D imaging methods \cite{levoy_light_2006,broxton_wave_2013,skocek_high-speed_2018,xue_single-shot_2020,boominathan_recent_2022}. 
These methods encode volumetric data into a single 2D image, which allow for computational reconstruction of the object. 
One prominent technique is the light field microscope \cite{levoy_light_2006,broxton_wave_2013,skocek_high-speed_2018}, which utilizes a standard microlens array to enable 3D capabilities. 
Recently, lensless architectures using coded masks or randomized microlens diffusers, have also been demonstrated~\cite{adams_single-frame_2017,antipa_diffusercam_2018,kuo_-chip_2020, xue_single-shot_2020,tian_geomscope_2021}. 

While these approaches are effective, they also come with associated limitations.
A significant challenge with these systems, with the exception of the Fourier light field microscope \cite{guo_fourier_2019,han_3d_2022}, is that their point spread functions (PSFs) are not laterally shift-invariant, which necessitates extensive calibration to define the PSF accurately and restricts measurement to the pre-calibrated volume. 
Further, the shift-variant nature of the PSF complicates the image analysis -- reconstructions typically require optimization algorithms that impose sparsity constraints.
Another drawback of using coded masks or microlens arrays is that the large size decreases the peak signal level of the PSF by spreading photons over a larger area. Additionally, photons from various axial and lateral positions overlapping on the sensor further degrade the signal-to-noise ratio (SNR). 
To address some of these issues while improving resolution and depth of field (DOF), Miniscope3D \cite{yanny_miniscope3d_2020} and Fourier DiffuserScope \cite{linda_liu_fourier_2020} replace the tube lens in a standard fluorescence microscope with an optimized phase mask, but they still suffer from some of the limitations associated with non-compact PSFs.

PSF engineering is an alternative approach that modulates the fluorescence at the back focal plane (BFP) of the objective lens in the imaging system to encode the axial position of the emitters within compact PSFs. 
Numerous 3D PSFs have been proposed over the past two decades \cite{von_diezmann_three-dimensional_2017,Nehme2020}, including the astigmatic PSF \cite{huang_three-dimensional_2008}, which varies in elongation direction and magnitude with defocus; the corkscrew \cite{lew_corkscrew_2011} and double-helix \cite{Pavani2009} PSFs, which feature revolving spots around the emitter; and the tetrapod \cite{Shechtman2015} and pixOL \cite{Wu2021.12.30.474544} PSFs, which are optimized for maximizing the Fisher information. 
Another group of PSFs, including the bisected \cite{Backer2014a}, quadrated \cite{Backer2013}, and tri-spot PSFs \cite{Zhang2018}, images subapertures off the pupil center to induce lateral displacements in the focal point; the displacement direction depends on the position of the subaperture relative to the center of the BFP, and the displacement amount is proportional to the defocus.

While these engineered PSFs have proven effective primarily in single molecule localization microscopy \cite{Hess2006,Rust2006,Betzig2006,lelek_single-molecule_2021} where isolated point sources are imaged, reconstructing more complex geometries from these PSFs remains challenging.
Unlike methods like Miniscope3D, compact PSFs usually cannot satisfy the multiplexing requirement of compressed sensing.
Therefore, recovering the 3D object using a single image often results in ambiguities in depth measurements \cite{ghanekar_ps_2024,zhang_investigating_2024}. 
Although using multiple images with engineered PSFs has the ability to substantially reduce these ambiguities, such as in the Fourier light field microscope \cite{guo_fourier_2019,han_3d_2022} and complex-field and fluorescence microscopy using the aperture scanning technique (CFAST) \cite{zhang_investigating_2024},
these methods require either capturing multiple images sequentially at the expense of temporal resolution or utilizing different areas of the detector to perform spatial multiplexing, thus sacrificing the FOV.
A recent approach, the polarized spiral PSF \cite{ghanekar_ps_2024}, integrates polarizers with a double-helix phase mask and employs orthogonally polarized detection channels from a polarization camera to achieve single-shot 3D imaging without sacrificing either the temporal resolution or the FOV. However, it has not completely eliminated the ambiguity problem.

Another critical aspect of SVF imaging is the reconstruction algorithm; a robust reconstruction algorithm is essential for accurately and precisely reconstructing 3D scenes from 2D measurements captured with engineered PSFs. 
The Richardson-Lucy (RL) algorithm \cite{guo_fourier_2019,kak_principles_2001,dellacqua_model-based_2007} is broadly used due to its effectiveness in recovering 3D structures. 
However, its performance degrades under noisy conditions, particularly in fluorescence imaging where the signal is limited.
Given these limitations, it is worth exploring the use of neural fields in SVF. The topic of neural field is a recent prominent and prevalent technique for 3D scene representations and graphics rendering \cite{Park2019DeepSDFLC, mildenhall2020nerf}. Neural field techniques perform mapping between spatial coordinates and image intensity values using a compact multi-layer perceptron (MLP) model and optionally learnable positional embeddings.
This approach has recently been applied to improve microscopic systems. 
For example, it has been implemented to represent continuous 3D refractive index in intensity diffraction tomography and to generate continuous images across the axial dimension \cite{liu2022decaf}, but it requires extensive optimization time on high-performance graphics processing units (GPU). 
Further developments reduced the computational resources by creating an additional hash encoding layer at the input of the neural network for 2D microscopic imaging systems \cite{xie2023diner}. 
More adaptations of the neural fields approach include a 2D version for lensless imaging phase retrieval \cite{Zhu:22} and the modeling of space-time dynamics with 4D neural fields in imaging through scattering~\cite{Feng2023NeuWSNW}, computational adaptive optics \cite{kang2024coordinatebased}, and structured illumination microscopy \cite{Cao2024SpaceTime}. 
Advancements continued with the exploitation of redundancies in Fourier ptychographic microscopes to augment the MLP model with a compact learnable feature space, thereby speeding up the image stack reconstructions and reducing data storage burdens \cite{Zhou2023fpminr}. 
Most recently, the integration of neural fields with diffusion models has been proposed to aid volumetric reconstruction \cite{hui2024microdiffusion}.

\begin{figure}[ht!]
    \centering
    \includegraphics{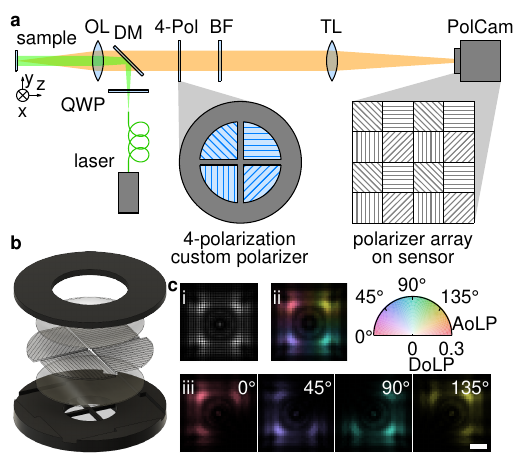}
    \caption{(a) Schematic of single-shot volumetric fluorescence imaging using the QuadraPol PSF. OL, objective lens; TL, tube lens; QWP, quarter waveplate; DM, dichroic mirror; BF, bandpass filter. A 4-polarization custom polarizer (4-Pol) is positioned at the back focal plane (BFP) of the imaging system to modulate the emission light, with a polarization camera (PolCam) capturing the modulated fluorescence. The transmission axes of the polarizer and PolCam are 0°, 45°, 90°, and 135°. (b) Assembling the custom polarizer by aligning two coverslips and four laser-cut polymer polarizers between 3D-printed holders. (c) A representative image of a point source captured by the polarization camera, visualized using (i) raw pixel readouts, (ii) a polarization image color-coded in the HSV scheme (AoLP as hue, DoLP as saturation, and intensity as value), and (iii) four separate images for each polarization channel. Scale bar: 50 \textmu{}m.}
    \label{fig:schematic}
\end{figure}

In this paper, we present QuadraPol PSF, an engineered PSF designed for SVF imaging that is easy to implement and overcomes the limitations of existing techniques.
Additionally, we introduce a reconstruction algorithm based on neural fields that outperforms the widely used RL deconvolution. 
Our work is inspired by the algorithm architectures from \cite{Feng2023NeuWSNW, Zhou2023fpminr,hui2024microdiffusion}, which we redesigned to enhance the capabilities of our imaging system, as demonstrated through fluorescence imaging over a centimeter-scale lateral FOV, with a depth of 5 mm, a lateral resolution of $\sim$7 \textmu{}m, and an axial resolution of $\sim$240 \textmu{}m. 
QuadraPol PSF enables all-in-focus imaging of bacterial colonies on sand surfaces and 3D visualization of plant root morphology. While our focus has been on the rhizosphere, the high-quality structural images produced suggest that this approach can be effectively generalized to other contexts as well.
Our results highlight the potential of QuadraPol PSF combined with neural fields to markedly propel biological research through enabling rapid, high-resolution 3D imaging of intricate biological structures across a large FOV.

\section{Methods}

\subsection{Imaging system}

\begin{figure*}[ht!]
    \centering
    \includegraphics{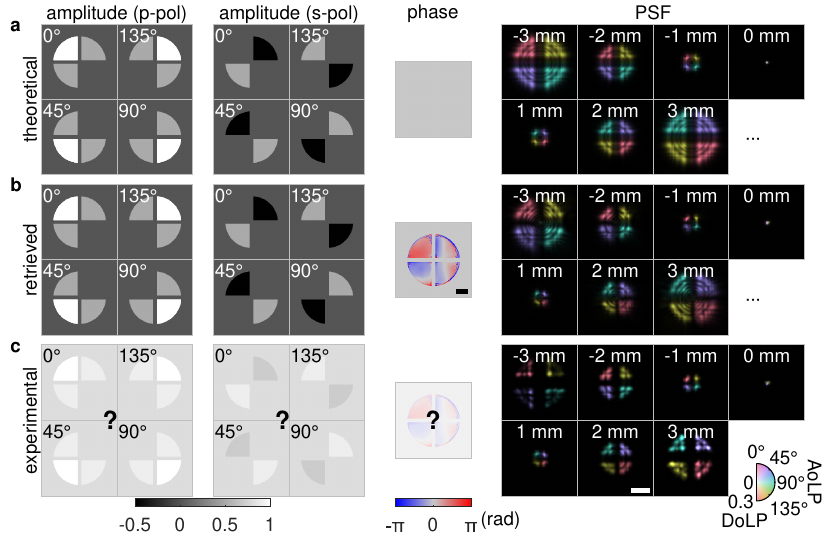}
    \caption{Amplitude and phase of the pupil and PSFs for at different heights. (a) Theoretical PSFs without aberration, (b) simulated PSFs using the retrieved phase, and (c) experimental PSFs. Question marks indicate that the phase and amplitude for the experimental PSF are not accessible. Scale bar: 2 mm for the pupil images, and 0.2 mm for the PSF images.
    }
    \label{fig:psf}
\end{figure*}

The experimental setup of our SVF imaging system with the QuadraPol PSF is shown in Figure \ref{fig:schematic}(a). 
We modulate a 568 nm laser (Coherent Sapphire 568 LP) transmitted through a single-mode fiber (Thorlabs P3-460B-FC-2) using a quarter-wave plate (QWP, Thorlabs WPQ10M-561) to produce circularly polarized excitation at the sample. 
An achromatic doublet lens (Thorlabs AC254-080-A) is used as the objective lens (OL), providing access to the back focal plane (BFP) without the need for additional 4f systems, thus simplifying the setup. 
The collected fluorescence is then filtered by a dichroic mirror (DM, Semrock Di01-R488/561) and a bandpass filter (BF, Semrock FF01-523/610), and modulated by a custom 4-polarization polarizer (4-Pol). 
The modulated fluorescence is then focused onto a polarization camera (The Imaging Source DZK 33UX250) using a tube lens (TL, Thorlabs AC254-150-A). 
This compact system is mounted on a $z$-stage (Newport 436, controlled by Newport CONEX-LTA-HS) and a custom-built $xy$ gantry system (Ballscrew SFU1605 with NEMA23 Stepper Motor) capable of scanning a 200$\times$200 mm$^2$ area.

The 4-Pol custom polarizer (Figure \ref{fig:schematic}(b)) is constructed by cutting four pieces from a thermoplastic polymer film linear polarizer (Thorlabs LPVISE2X2) with a laser cutter (Universal Laser Systems PLS6.75), and positioning them between two circular coverslips (VWR 48380-046). 
Two 3D-printed (Anycubic Photon Mono X 6Ks) holders apply clamping forces to secure the polarizers and coverslips, preventing displacement. One holder features a 9-mm diameter aperture to ensure the shift invariance of the imaging system’s PSF, corresponding to a numerical aperture (NA) of 0.056.
Additionally, 0.6-mm lines are printed along the center of the holder to block unmodulated fluorescence from passing through gaps between the polarizers.

The polarization camera integrates a polarizer microarray atop the CMOS sensor. 
Typically, the raw pixel data (Figure \ref{fig:schematic}(c-i)) is processed to visualize the angle of linear polarization (AoLP), which indicates the polarization direction and the degree of linear polarization (DoLP), which quantifies the proportion of polarized fluorescence (Figure \ref{fig:schematic}(c-ii)) \cite{born2013principles,bruggeman_polcam_2023}. Alternatively, the data can be decomposed into images corresponding to the four polarization channels, as shown in Figure \ref{fig:schematic}(c-iii).

\subsection{Point spread function design and experimental calibration}

The design of the QuadraPol PSF is inspired by the quadrated PSF \cite{Backer2013}, the multi-view reflector (MVR) microscope \cite{zhang_six-dimensional_2023}, and CFAST \cite{zhang_investigating_2024}. 
These techniques divide the pupil into four sections. 
Specifically, the quadrated phase mask applies different phase ramps to each segment, generating a four-spot PSF in a single imaging channel. 
In contrast, the MVR and CFAST systems utilize reflective mirrors and light-blocking apertures, respectively, to create distinct image channels for each pupil segment. 
A key characteristic of these techniques is the encoding of the axial position of the emitter through the different lateral displacements of the focused spots, each associated with one of the four equal segments of the pupil. 
However, these approaches come with limitations. The single-image capture with the quadrated PSF can lead to ambiguities in 3D reconstruction, while the MVR and CFAST setups sacrifice either FOV or temporal resolution. 
The QuadraPol PSF distinguishes itself by modulating each pupil section using a polarizer with a unique transmission axis, allowing the polarization camera to simultaneously capture four imaging channels without losing FOV or temporal resolution.

\begin{figure}[t!]
    \centering
    \includegraphics[width=3.45in]{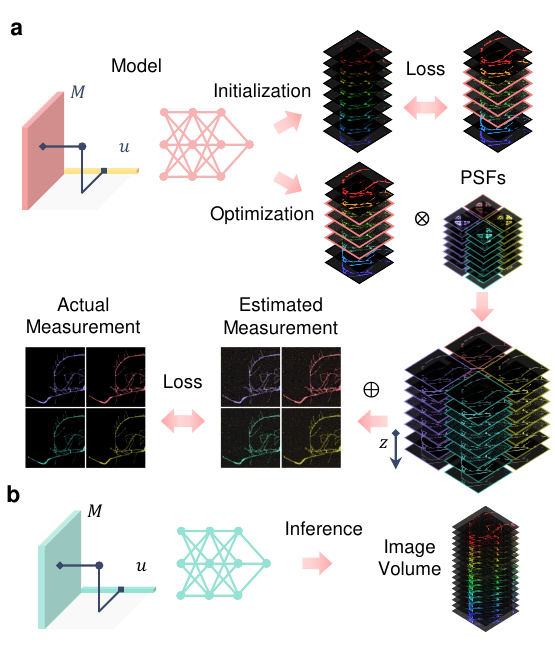}
    \caption{Framework of using neural fields to extend the quality and depth range of the imaging system. (a) The RL-deconvolved image volume guides the initialization of the model with a compact learnable feature space and multilayer perceptron (MLP). After model initialization, the model is further optimized for the image volume. The estimated image volume goes through the forward model of the imaging system to generate the estimated measurements. These measurements are compared with the acquired measurements, and then to update the model weights and parameters. (b) Once the model is optimized, the parameters and weights are fixed. It can render an image stack with continuous sampling. The operator $\otimes$ denotes a convolution operation, and $\oplus$ indicates a summation operation along $z$-axis.}
    \label{fig:neural}
\end{figure}
Given the imaging system’s low NA, the theoretical PSF (Figure \ref{fig:psf}(a)) for the axial position $z$ in each polarization channel ($\mathrm{pol}\in\{0^\circ,45^\circ,90^\circ,135^\circ\}$) can be described by (see Supplementary Note S1 for more details) \cite{novotny2012principles,Backer2014}
\begin{align}
    \mathrm{PSF}_{z,\mathrm{pol}}=\sum_{i\in[p,s]}\mathcal{F}\{A_{\mathrm{pol},i}(u,v)\exp\{j[kz&\sqrt{1-u^2-v^2}\nonumber \\ &\hspace{12pt}+P(u,v)\big]\big\}\big\}^2,
\end{align}
where $k = 2\pi / \lambda$ represents the wave number, and $(u, v)$ are the coordinates at the BFP. The pupil phase is consistent across all polarization channels; a uniform phase, $P(u, v) = 0$, is used for calculating the unaberrated theoretical PSF. 
The amplitude modulation patterns for $p$- and $s$-polarizations are shown in Figure \ref{fig:psf}(a). 
These patterns assign values of 1, 0.5, and 0 for amplitude mask with $p$-polarization, and 0, $\pm$0.5, and 0 for $s$-polarization, corresponding to the angles between the transmission axis of the polarizers and detection channel at 0°, 45°, and 90°, respectively.
Note that due to cross-talk between polarization channels separated by 45°, the PSF in each channel exhibits a primary spot and two weaker side spots, rather than a singular focal point (Supplementary Figure S1).
These images exhibit apparent lateral shifts in different directions across different polarization channels as the object defocuses; this is analogous to viewing the same object from four unique perspectives, which fundamentally enables 3D reconstruction.

Given the presence of aberrations in our imaging system, primarily introduced by the custom-made polarizer, we use the vectorial implementation of phase retrieval \cite{Ferdman2020} to compensate for these imperfections. 
This approach enables us to refine a series of retrieved PSFs (Figure \ref{fig:psf}(b)) that more closely match those obtained experimentally (Figure \ref{fig:psf}(c)) compared with theoretically obtained ones (Figure \ref{fig:psf}(a)). 
These experimental PSFs are generated by capturing images from fluorescent beads (Thermo Fisher F8858) that were axially scanned over an 8 mm range with a 0.1 mm step size. 
The retrieved phase of the pupil is shown in Figure \ref{fig:psf}(b), while we assume that the pupil amplitude remains the same as those in Figure \ref{fig:psf}(a). 
Both experimental and retrieved PSFs are implemented in the image reconstruction in Section \ref{sec:root}.

\subsection{Reconstruction algorithm}\label{sec:inr}

Previous studies have demonstrated the effectiveness of the modified RL deconvolution algorithm in reconstructing 3D volumetric scenes using multiple input images \cite{guo_fourier_2019,zhang_investigating_2024};
the algorithm iteratively updates the estimated object $o$ as
\begin{equation}
    o^{(k+1)} = o^{(k)} \frac{\sum_{\mathrm{pol}}\left(I_{\mathrm{pol}}\otimes\mathrm{PSF}_{z,\mathrm{pol}}^*\right)}{\sum_{\mathrm{pol}}\left\{\left[\sum_{z}\left(o_z^{(k)}\otimes\mathrm{PSF}_{z,\mathrm{pol}}\right)\right]\otimes\mathrm{PSF}_{z,\mathrm{pol}}^*\right\}},
\end{equation}
where $\mathrm{PSF}^*$ represents the original PSF with a 180° rotation in the $xy$ plane, $I_{\mathrm{pol}}$ represents the intensity from one of the polarization channels, and $\otimes$ denotes the convolution operator.

One limitation of the RL deconvolution is its performance under noisy situations.
Inspired by the forward imaging model, which captures four independent perspectives of the 3D scene,
we customized and designed the neural fields to enhance the performance of our system.
The algorithm is composed of a two-stage optimization process.
The first stage was to initialize our neural fields with the RL-deconvolved image volume. 
The neural field learns a compact learnable feature space consisting of a 3D feature tensor ($M$) and a feature tensor ($u$) \cite{Zhou2023fpminr}, and it also learns an MLP network including two nonlinear layers with ReLU activation functions and one linear layer. 
Each layer of the MLP in our neural field has $Q$ neurons with an additional offset neuron. 
The number of neurons is designed to match the number of feature channels in the feature space. 
The feature space is created from the Hadamard product of a feature tensor $M$ with a size of $N/2 \times N/2 \times Q$ and a feature tensor $u$ with a size of $Z\times Q$, where $N$ is the number of pixels along $x$ or $y$ axis of the raw image, $Q$ is the number of feature channels, and $Z$ denotes the number of pre-defined $z$-coordinates. 
The detailed parameters and hyperparameters of the neural fields can be found in Supplementary Note S2 \cite{loshchilov2019decoupledweightdecayregularization,fu2019retinamasklearningpredictmasks}.
The optimization process seeks to find a mapping function ($\phi$) between feature tensor space and the image volume
\begin{equation}
    \phi(\bv{x}): \mathbb{R}^2\cdot \mathbb{R} \mapsto \mathbb{R}^3, \;\; \mathbb{C} (f(\bv{x}), \phi(\bv{x}) ),     
\end{equation}
where $\bv{x}=(x,y,z)$ represents the 3D coordinate system and $\mathbb{C}$ is a set of constraints bounded by the image volume $f(\bv{x})$. 
This initialization process can significantly shorten the rendering time.
After the first stage, the neural field can render an image volume as an initial guess ($g$).

At the second stage, we can render our estimated measurements
\begin{equation}
    I_\mathrm{pol}(x,y) = \sum_{z} | g_{z}(\bv{x}) \otimes \mathrm{PSF}_{z,\mathrm{pol}} |^2,
\end{equation}
where $ I_\mathrm{pol}(x,y) $ is the estimated measurements from the neural field rendered image volume, $g_{z}$ denotes the rendered image volume at one $z$-plane, and $\otimes$ is the convolution operator. 
At our second stage of optimization, we optimized the neural field to minimize the difference between the estimated measurements and the experimental intensity measurements (Figure \ref{fig:neural}(a)). 

\begin{figure*}[ht!]
    \centering
    \includegraphics{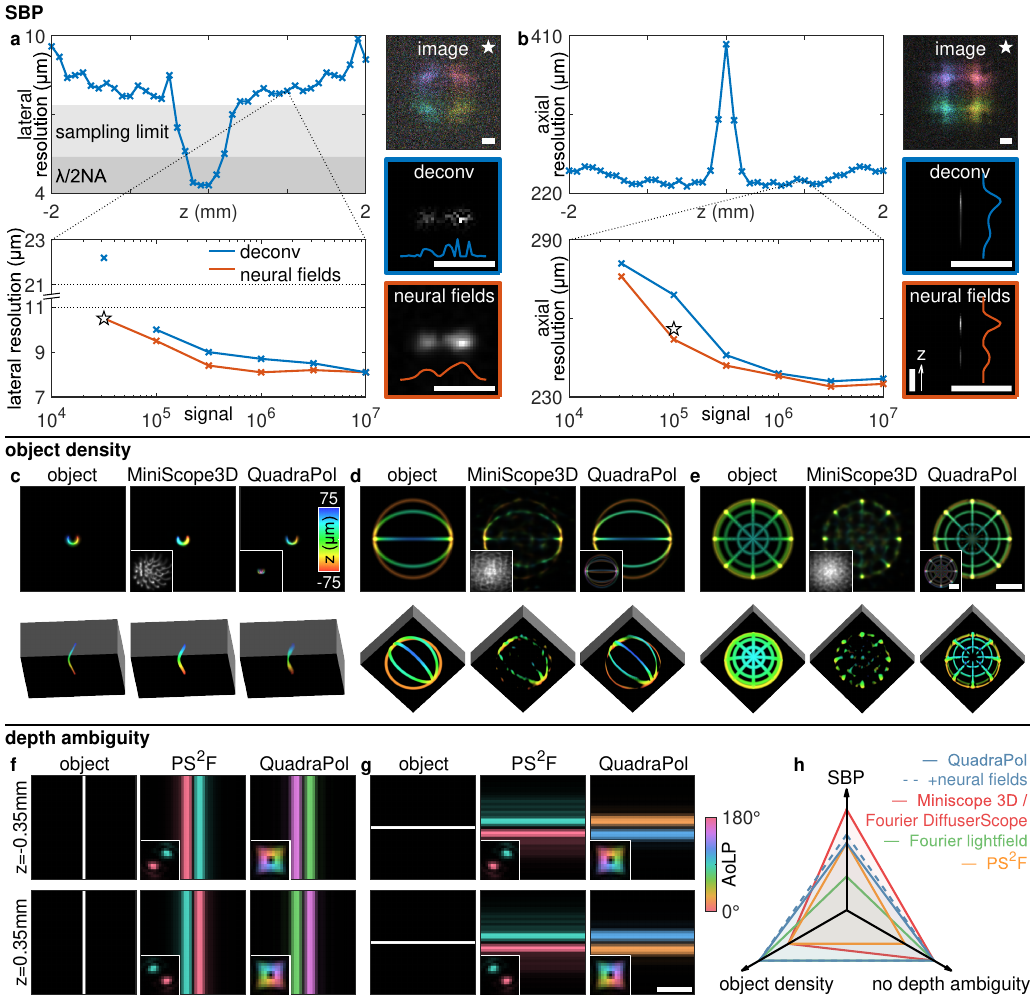}
    \caption{Performance evaluation of the QuadraPol PSF using simulated data. (a) Lateral and (b) axial resolutions determined by the Rayleigh criterion as functions of axial position and signal level. Images show representative data with Poisson shot noise and reconstruction cross-sections using RL deconvolution and neural fields. Shaded areas represents the diffraction ($\lambda/2\mathrm{NA}$) and sampling limits. Scale bar: 20 \textmu{}m in $xy$ view and 200 \textmu{}m in $xz$ view. (c-e) Reconstruction results for various line structures; insets show simulated images. Parameters for simulating the QuadraPol PSF ($\mathrm{NA}=0.52$, magnification $M=-5.2$, camera pixel size $2.2$ \textmu{}m) are adjusted to match those reported for Miniscope3D \cite{yanny_miniscope3d_2020}. Scale bar: 100 \textmu{}m; color bar: height in \textmu{}m. (f,g) Object and images using the polarized spiral (PS$^2$F) and QuadraPol PSFs with (f) vertical and (g) horizontal lines at $z=\pm0.35$ mm. Scale bar: 50 \textmu{}m; color bar: AoLP. Insets show the PSFs. (h) Comparison between the QuadraPol PSF and other SVF techniques.}
    \label{fig:comparison}
\end{figure*}

After the two-stage optimization, the parameters and weights in the neural field are fixed. 
We can continuously sample the feature tensor ($u$) to render a denser image volume along the axial ($z$) axis, as shown in Figure \ref{fig:neural}(b). 
Throughout the whole reconstruction process, no pre-training or training datasets are needed, and the algorithm is only supervised by a single capture.
The average optimization time over 196 FOVs in the plant root experiment (see Section 3.4 for more details) on an Nvidia A100-80GB GPU device is $\sim$22 seconds.

\subsection{Sample preparation}

Starting from a single colony, \textit{E. coli}-mScarlet-I was grown in LB medium overnight at 37°C. 
On the next day, the overnight culture was washed once in minimal medium and diluted to an optical density of 1. 
To prepare for imaging, 1 mL of cell culture was added to 10 g of autoclaved fine sand (Fischer Science Education Sand, Cat. No. S04286-8) in a small petri dish and briefly mixed by vortexing before imaging.

Soft white wheat seeds were obtained from Handy Pantry (Lot 190698). 
Seeds were sterilized by incubating in 70\% ethanol for 2 min, washing 3 times with autoclaved water, incubating with a bleach (50\% v/v) and Triton X (0.1\% v/v) solution for 3 min, and finally washing 5 times with autoclaved water. 
Seeds were then plated on 0.6\% phytagel (Sigma Aldrich, Cat. No. P8169) containing 0.5x MS medium (Sigma Aldrich Cat. No. M5519) and transferred to a growth chamber with a day/night cycle of 16 hr/8 hr at 25°C for 7 days. 
On day 7, seedlings were collected and incubated in 100 \textmu{}M merocyanine 540 (MC540) in phosphate buffered saline (PBS) for 15 min, then rinsed in DI water before imaging.

\section{Results}

\subsection{Simulation evaluation of system performance}
\label{sec:sim}

We first assessed the resolution of the QuadraPol PSF by simulating images of two closely positioned point sources.
To determine the resolution, we applyed the Rayleigh criterion, which requires at least a 20\% dip between the peaks of the two reconstructed spots.
We used a high spatial sampling rate to evaluate the resolution more accurately.
In the lateral direction, the QuadraPol PSF achieves an in-focus resolution of 4.3 \textmu m.
However, this resolution rapidly degrades to approximately 7 \textmu m with a $\pm0.3$ mm defocus.
The resolution remains under 9 \textmu m with a defocus extending up to 1.8 mm (Figure \ref{fig:comparison}(a)).
To optimally adapt to these resolutions in our experimental setup, we used a camera sampling rate of 3.7 \textmu m in the object space (with a magnification of 1.875) that corresponds to a resolution limit of 7.4 \textmu m, achieving a DOF of $\sim$4 mm with relatively uniform lateral resolution. 
This extended DOF is notably longer than the 0.26 mm DOF achievable with a standard PSF, which is due to the smaller effective NA of the QuadraPol PSF.

Additionally, we compared our neural fields reconstruction algorithm with the RL deconvolution. 
While both algorithms perform similarly under ideal conditions, neural fields exhibit significantly improved robustness with noisy images (Figure \ref{fig:comparison}(a) and Supplementary Figure S2).
For instance, at a low signal level where a total of $3.16\times10^{4}$ photons are captured from both emitters, the RL deconvolution results in a highly noisy reconstruction, whereas the neural fields can still distinctly resolve the two spots and offer a two-fold resolution improvement. 
{This improvement is attributed to the fact that the neural fields tend to favor smooth reconstructions~\cite{mildenhall2020nerf, tancik2020fourier, Feng2023NeuWSNW, Cao2024SpaceTime}, effectively leading to a suppression of noise on the reconstruction process.}

The axial resolution within the 4 mm DOF ranges from 230 to 250 \textmu{}m, except for positions close to the focal plane ($|z|<0.1$ mm), where it degrades to 405 \textmu{}m (Figure \ref{fig:comparison}(b)).
This degradation is attributed to the symmetry which causes gradual depth gradient of the in-focus QuadraPol PSF; both $+z$ and $-z$ defocus result in an expansion of the PSF.
However, this symmetry is disrupted in the presence of aberrations; when aberrated PSFs were simulated, an improvement in axial resolution near the focal plane was observed (Supplementary Figure S3).
We also experimentally demonstrated this improvement in Section \ref{sec:beads}.
Similar to the case of lateral direction, using neural fields leads to a better axial resolution compared to RL deconvolution under noisy conditions, although the improvement is less pronounced than in the lateral direction, particularly at very low signal levels. This is likely due to the design of the network's feature space, which provides more degrees of freedom in the lateral direction than in the axial direction, given the lower axial resolution of the imaging system compared to its lateral resolution. While such a configuration can significantly accelerate the algorithm and reduce memory usage, it can also limit the improvements in axial resolution.
Again, we emphasize that this comparison is under ideal conditions and does not account for forward model mismatches that may occur in experimental settings.
The space-bandwidth product (SBP) of our system is $\sim$5.2 million voxels over the 3.8 mm $\times$ 4.5 mm FOV and 4 mm depth range.
Note that the SBP for the biological experiments in later sections is greater than the value reported here. We are not confined to the 4 mm depth range, as we can still reconstruct the object with slightly reduced resolution at greater defocus distances.

One key advantage of the QuadraPol PSF is its ability to resolve denser objects more effectively.
By separating polarization channels, the compact QuadraPol PSF reduces the mixing of information from different 3D positions, thus relaxing the sparsity constraint in the reconstruction algorithm compared to other SVF techniques such as the Miniscope3D \cite{yanny_miniscope3d_2020}.
To demonstrate this, we simulated three different line structures and generated images using both the Miniscope3D PSF and the QuadraPol PSF, analyzing them with the same RL deconvolution algorithm without sparsity constraints.
For a small structure where the raw image using the Miniscope3D shows minimal overlap, both methods accurately resolve the structure (Figure \ref{fig:comparison}(c)).
Additionally, the reconstruction quality of the Miniscope3D for defocused axial positions is superior due to its extended DOF.
However, for larger objects where there is substantial information mixture in the raw Miniscope3D images (Figure \ref{fig:comparison}(d,e)), it struggles to accurately recover the structure.
In contrast, the QuadraPol PSF still resolves the object effectively with significantly fewer artifacts.
This enhanced performance is similarly observed with other dense objects (Supplementary Note S3).

Another challenge in SVF imaging is depth ambiguity.
Compact PSFs, such as the double-helix PSF \cite{Pavani2009}, often produce identical images for specific structures at different heights (Supplementary Figure S4), which makes it impossible to accurately estimate depth without error. 
The polarized spiral PSF \cite{ghanekar_ps_2024} was developed to address this issue by integrating orthogonally-polarized polarizers on the left and right sides of a double-helix phase mask, creating two spots with orthogonal polarizations captured using a polarization camera. 
While this method resolves the ambiguity for vertical line structures (Figure \ref{fig:comparison}(f)), challenges remain for horizontal lines (Figure \ref{fig:comparison}(g)). 
Clearly, to eliminate ambiguity in all directions, at least three independent images are necessary, highlighting the need to use the full capability of the polarization camera. 
As shown in Figure \ref{fig:comparison}(f,g), the QuadraPol PSF effectively solves this issue, allowing both horizontal and vertical lines at different depths to be resolved without ambiguity.

\begin{figure}[b!]
    \centering
    \includegraphics{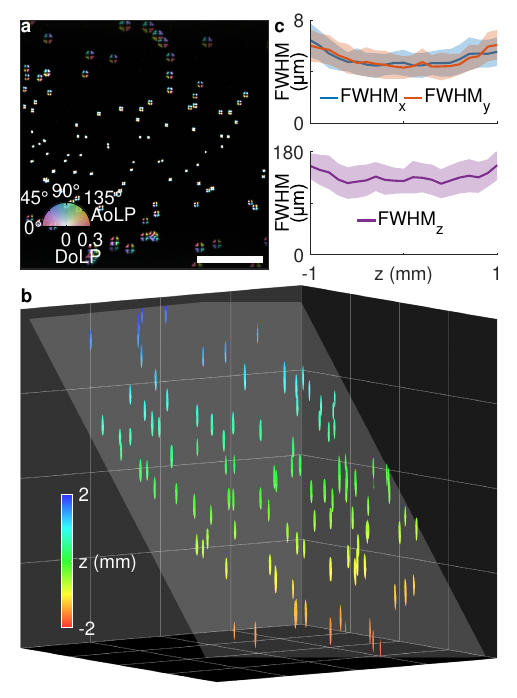}
    \caption{Imaging fluorescent beads on a 45°-tilt coverslip using the QuadraPol PSF. (a) Raw image of the fluorescent beads. Scale bar: 1 mm. (b) Three-dimensional rendering of the reconstructed beads using MATLAB function ``isosurface''. Grid size: 1 mm. (c) Full width at half maximum (FWHM) values for the reconstructed beads. Lines represent the average; shaded areas represent the standard deviation.}
    \label{fig:beads}
\end{figure}

A comparison of the three aspects mentioned above between the QuadraPol PSF and other SVF techniques is shown in Figure \ref{fig:comparison}(h). 
The most significant advantage of the QuadraPol PSF, which we would like to emphasize, is its ability to eliminate depth ambiguity and reconstruct 3D scenes without relying on sparsity constraints, distinguishing it from other SVF techniques such as MiniScope3D, Fourier DiffuserScope, and PS$^2$F.
Additionally, compared to the Fourier light field microscope, which shares the aforementioned advantages, the QuadraPol PSF achieves a higher SBP as it does not compromise the field of view (FOV).
Furthermore, the QuadraPol PSF provides a better SNR, despite the photon loss due to the linear polarizers (Supplementary Note S3).

\subsection{Experimental validation using fluorescent beads}
\label{sec:beads}

We next validated the QuadraPol PSF with fluorescent beads placed on a tilted coverslip (Figure \ref{fig:beads}(a)). 
The 3D reconstruction is shown in Figure \ref{fig:beads}(b), where the reconstruction algorithm accurately resolves the fluorescent beads on the coverslip without noticeable inaccuracies. 
Quantitative analysis of depth accuracy is provided in Supplementary Figure S5.
Additionally, we quantified the full width at half maximum (FWHM) values for the reconstructed images of 72 fluorescent beads from a $z$ scan (Figure \ref{fig:beads}(c)). 
The FWHM values are $4.7 \pm 1.0$ \textmu m and $4.3 \pm 0.9$ \textmu m in the $x$ and $y$ directions, respectively, for in-focus beads. 
The average FWHM remains within 5.0 \textmu m for a $|z| \leq 0.6$ mm. When beads are defocused by 1 mm, the FWHM increases to an average of 6.0 \textmu m.
Note that although the FWHM is not directly comparable to the resolution determined using Rayleigh criteria in Section \ref{sec:sim}, the trend observed as a function of $z$ is consistent with our simulations (Figure \ref{fig:comparison}(a)).
In the $z$ direction, the average FWHM is consistently below 130.4 \textmu m across a DOF of 1.2 mm and increases to an average of 155.7 \textmu m when defocused by 1 mm. 
We do not observe resolution degradation near the focal plane, unlike what was seen in simulations using theoretical PSFs (Figure \ref{fig:comparison}(b)). This reaffirms that the system's aberrations actually improve the depth resolution of the QuadraPol PSF. Additional resolution analysis (Supplementary Figure S6) based on the Rayleigh criterion using synthesized images from bead data is consistent with the simulation results shown in Figures \ref{fig:comparison}(a,b).

\subsection{All-in-focus imaging of bacterial colony on sand surfaces}

\begin{figure*}[t!]
    \begin{center}
    \includegraphics{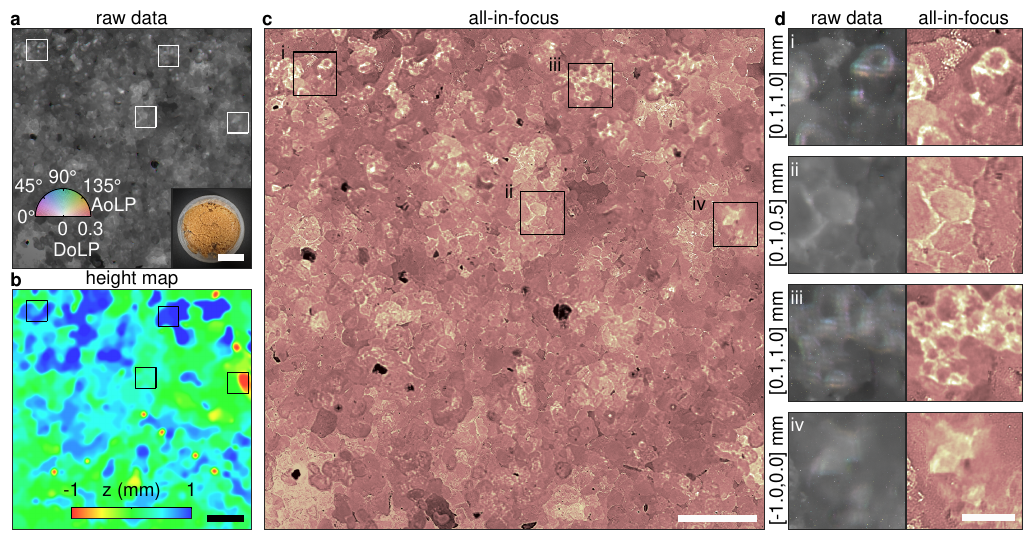}
    \caption{All-in-focus imaging of \textit{E. coli} tagged with mScarlet-I on sand surfaces. (a) Raw polarized fluorescence image. The inset shows a photograph of the sample captured using a smartphone camera. (b) Height map recovered using the all-in-focus algorithm. Color bar: height in mm. (c) The all-in-focus fluorescence image. (d) The zoomed regions of interest in (a-c). Annotations represent the $z$ range for each region. Scale bar: 20 mm in the inset of (a), 2 mm in (b,c), and 0.5 mm in (d).}
    \label{fig:sand}
    \end{center}
\end{figure*}

Bacterial activities play a crucial role in the biochemical processes within the rhizosphere \cite{fujishige_feeling_2006}.
Given the highly complex and scattering nature of soil environments, many lab-based environmental and biological experiments use sand as a proxy \cite{vollsnes_quantifying_2010,probandt_microbial_2018,anselmucci_imaging_2021}.
Sand offers better physical and chemical simplicity and uniformity, providing a well-controlled setting for scientific analyses. 
Visualizing bacterial colonies in sand represents a significant advancement toward \textit{in situ} studies of bacterial behavior. 
However, a key challenge arises from the typical size of sand particles, which creates a non-flat surface that can cause many areas within the FOV to be out of focus when captured in a single snapshot by a standard microscope without a $z$ scan.

To address this challenge, we use 3D imaging with the QuadraPol PSF to visualize \textit{E. coli} tagged with mScarlet-I on the sand surface (Figure \ref{fig:sand}(a)). 
Multiple FOVs are stitched together using the Microscopy Image Stitching Tool plugin in imageJ \cite{chalfoun_mist_2017}. 
To determine the correct focusing height map (Figure \ref{fig:sand}(b)) and generate an all-in-focus image (Figure \ref{fig:sand}(c)) \cite{bian_autofocusing_2020,liang_all--focus_2022}, we use a $100\times100$ pixel window (side length of 0.35 mm) sliding across the entire FOV. 
The sharpest focus $z$ position for each window is selected based on image contrast, defined as the difference between the 99th and 1st percentile values in each slice in the reconstructed $z$-stack (Supplementary Figure S7). 
The zoomed regions (Figure \ref{fig:sand}(d)) show that we successfully resolve sharp reconstructions of the bacterial colonies on the sand particles from blurred raw images across the entire sample area, which has a depth variation of 2 mm.

\subsection{Three-dimensional imaging of plant roots with neural fields}

\label{sec:root}

\begin{figure*}[t!]
    \centering
    \includegraphics{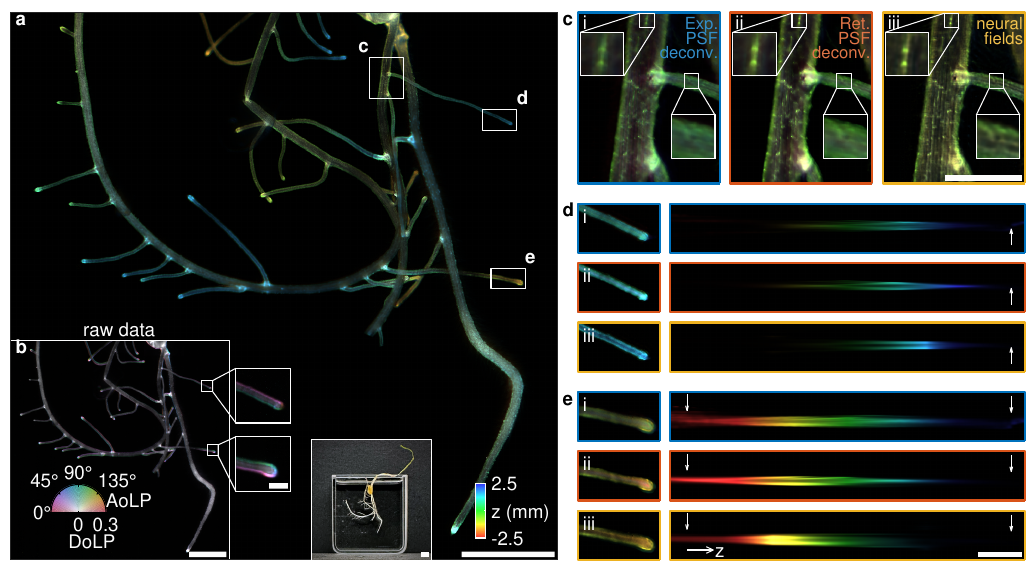}
    \caption{Volumetric imaging of wheat roots using the QuadraPol PSF. (a) Reconstruction using neural fields. Inset shows a photograph captured using a smartphone camera. (b) The raw polarized fluorescence image. (c) The $xy$ view, and (d,e) the $xy$ and $xz$ views of zoomed regions in (a), reconstructed using (i) deconvolution with experimental PSF, (ii) deconvolution with retrieved PSF, and (iii) neural fields. Scale bar: 5 mm in (a,b), 0.5 mm in the zoomed region of (b), and 1 mm in (c-e). Color bar: depth in mm. Note that the color in the $xy$-view is composed of the entire reconstructed volume, reflecting the intensity contribution from all $z$ slices; this view predominantly displays a color corresponding to the central height of the root.}
    \label{fig:root}
\end{figure*}

Plant roots play a critical role in the rhizosphere, serving as the primary interface between the plant and the soil, and significantly influencing the activities of soil microbes. 
However, visualizing roots poses significant challenges due to their typically large depth variation, which complicates the acquisition of focused images across the entire volume of interest. 
Here, we conduct 3D imaging of wheat roots stained with MC540 within a 5 mm-thick glass container (Figure \ref{fig:root}(a), Starna Cells 93-G-5). 
The raw image (Figure \ref{fig:root}(b)) shows how depth is encoded in the polarization of detected fluorescence. 
Two zoomed regions show different polarization characteristics: one shows the top of the root polarized at 0 degrees and the bottom at 90 degrees while the other exhibits the opposite pattern, indicating different defocus directions for these root segments.

We performed the reconstruction using the RL deconvolution with both the experimental and retrieved PSFs, as well as our proposed method based on neural fields in Section \ref{sec:inr}. 
For the neural fields method, we used experimental PSF for $|z|\leq2$ mm and retrieved PSF otherwise during the two-step optimization process (Supplementary Note S4). 
This is due to the degraded quality of experimental PSFs from the low SNR at large defocus distances.
In contrast, using both PSFs in a single deconvolution results in sharp discontinuities at the $z$ position transitioning between experimental and retrieved PSFs (Supplementary Note S4).
The image volumes with color-coded depth are presented in Figure \ref{fig:root}(a). 
Observations from full FOV images (Supplementary Figure S8) highlighted that for thicker segments of the roots, the deconvolution results with the experimental PSF appear noisier compared to the method using neural fields.
Meanwhile, deconvolution with the retrieved PSF shows a broader spread in $z$ (Supplementary Figure S9), due to minor mismatches between the experimental and retrieved PSFs that reduce $z$-axis accuracy and precision. 
The neural field method addresses these issues by maintaining the accuracy and precision of the deconvolution with the experimental PSF while avoiding its SNR problems.

We further examine the zoomed region of a thicker part of the root (Figure \ref{fig:root}(c)), where the neural field reconstruction (Figure \ref{fig:root}(c-iii)) shows significantly sharper cell walls compared to the deconvolution results with either the experimental (Figure \ref{fig:root}(c-i)) or the retrieved PSF (Figure \ref{fig:root}(c-ii)). 
This improvement is attributed to our neural field method jointly taking the contribution from fluorescence signals at extended depth regions and accurate PSFs of the experimental calibrations. 
The enhancements in both the lateral and axial directions are also observed in out-of-focus thinner regions of the root (Figures \ref{fig:root}(d,e)), where $xy$ views using the neural field method show sharper images and the $yz$ views exhibit a narrower spread compared to the deconvolution results using the retrieved PSF; the width mirrors the precision of the experimental PSF while overcoming its limitations of low SNR at larger defocus.
Improved reconstruction quality with neural fields is also demonstrated in additional zoomed regions (Supplementary Figures S10 and S11), showing mitigation of artifacts from deconvolution with the experimental PSF, and visualization of fine structures like root hairs, which cannot be resolved using deconvolution with the retrieved PSF.

\section{Discussion and conclusion}

Our proposed method combines the QuadraPol PSF with neural fields to achieve SVF imaging.
A major innovation in hardware is the usage of the four-polarization custom polarizer, which simultaneously encodes depth information across four images (three of which are independent) captured by a polarization camera.
This setup maximizes the capabilities of the polarization camera, completely eliminating the estimation ambiguity while maintaining a compact footprint, which allows the QuadraPol PSF to enable SVF without a sparsity constraint and thus achieve a better SNR compared to the current state-of-the-art SVF techniques.
With a single FOV of 3.8 mm $\times$ 4.5 mm, we experimentally demonstrate the ability to capture a $\sim$100 mm$^3$ cubic volume with lateral and axial resolutions of approximately 7 \textmu{}m and 240 \textmu{}m, respectively. 
The extended DOF offered by this system reduces acquisition time by approximately 20 times compared to traditional $z$-scan methods for capturing the same volume. 
The effectiveness of the QuadraPol PSF is shown in our all-in-focus imaging of bacterial colonies on sand surfaces, where it captures sharp images of most of the colonies in a single snapshot.

For the reconstruction algorithm, neural fields demonstrate better performance compared to deconvolution,
as demonstrated in our plant root imaging experiments where neural fields produce sharper images.
The neural field approach also offers additional advantages such as creating a continuous representation of the object and significantly decreasing data storage requirements by an order of magnitude. 
For instance, storing plant root data as discrete images with 81 axial slices requires approximately 62~GB, whereas neural fields of the same data requires only 6.12 GB with the same data precision.
Furthermore, this physics-based method is free from any pertaining or any dataset collection. It can also be easily adapted to different system settings.

SVF imaging using the QuadraPol PSF and neural fields provides a powerful tool for large-FOV imaging at high spatial-temporal resolutions.
The design principle can be easily adapted to meet specific application requirements. 
For instance, our current setup prioritizes high temporal resolution by using a polarization camera with a large effective pixel size, thus sacrificing spatial resolution. 
However, systems requiring a higher spatial sampling rate from the detector could use a standard camera combined with polarization optical elements and temporal multiplexing to address this limitation.
One current limitation of the QuadraPol PSF is that it relies on the expansion of four simultaneously-in-focus spots to estimate depth, which inherently limits the imaging depth. 
If the source-to-focal-plane distance increases, the PSFs in all four polarization channels expand, causing photons to spread across a larger area and consequently reducing the SNR at large defocus distances. 
In other words, due to the limited DOF extension of the QuadraPol PSF, its SBP still has room for improvement compared with the MiniScope3D \cite{yanny_miniscope3d_2020} or the Fourier DiffuserScope \cite{linda_liu_fourier_2020} (Figure \ref{fig:comparison}(h)).
Integrating the QuadraPol PSF with DOF extension methods used in these other systems, such as using an engineered metasurface to create PSFs in four different polarization channels with distinct focal planes, should improve the SBP. 
Further, these methods can potentially overcome the limitation of low depth resolution near the focal plane without introducing aberrations.
We anticipate that these sophisticated improvements of QuadraPol would be worth implementing as this technology matures.
Given the presence of a polarization camera in the QuadraPol PSF, a natural extension of the method would be to simultaneously measure polarization \cite{shen_monocular_2023}. Similar to what has been demonstrated with the MVR microscope \cite{zhang_six-dimensional_2023}, the QuadraPol PSF should exhibit high polarization sensitivity. Previous work in polarization imaging, such as 3D Mueller matrix imaging \cite{ushenko_3d_2021,sieryi_optical_2022,ushenko_insights_2024} and Stokes-correlometry \cite{ushenko_stokes-correlometry_2019}, has established frameworks for quantitative analysis of light polarization. This direction holds particular promise, as fluorescence polarization has been shown to be critical in plant studies \cite{verbelen_polarization_2000,baskin_disorganization_2004} and other areas of biological research \cite{brasselet_polarization_2023}.

In summary, our experimental results highlight the potential of SVF imaging using the Quadra-Pol PSF and neural fields, particularly for applications such as studying microbial interactions within the rhizosphere.
Additionally, we expect that our SVF system can be applied across various fields, including clinical and biomedical research. For instance, we tested our method on synthetic lymph node vasculature data \cite{jafarnejad_quantification_2019} (Supplementary Figure S12), experimental mouse kidney data (Supplementary Figure S13), and demonstration of 3D particle image velocimetry (Supplementary Figure S14 and Supplementary Video S1). The simplicity of the optical setup also holds the potential for miniaturization, which would be useful for \textit{in vivo} imaging in freely behaving animals. Moreover, in terms of technological advancement, our system introduces a novel approach to SVF design: instead of mapping 3D objects to 2D images, our method maps 3D objects to 3D measurements, with polarization serving as the third dimension. We anticipate this innovative approach will inspire future developments in SVF imaging system designs.

\begin{backmatter}
\bmsection{Funding}
The work reported in this paper is supported by the Resnick Sustainability Institute and the Heritage Medical Research Institute (HMRI-15-09-01) at Caltech. 

\bmsection{Acknowledgments}
We thank Tara Chari for constructing the \textit{E. coli} strain harboring mScarlet-I. We also thank Daniel Wagenaar from the Caltech Neurotechnology lab for the assistance in fabricating the custom polarizer and Panlang Lyu for the helpful discussions.

\bmsection{Disclosures}
The authors declare no conflicts of interest.

\end{backmatter}

\bibliography{references}

\end{document}